\documentclass[10pt]{article}
\usepackage{epsfig}

\begin{document}

\title{On some geometric features of the Kramer interior solution 
 for a rotating perfect fluid}
\author{
F.J. Chinea and M.J. Pareja \\        
Dept. de F\'{\i}sica Te\'orica II,  Ciencias F\'{\i}sicas, \\
Universidad Complutense de Madrid \\
E-28040 Madrid, Spain}
\date{}
\maketitle

\begin{abstract}
 Geometric features (including convexity properties) 
of an exact interior gravitational field due to a self-gravitating axisymmetric 
body of perfect fluid in stationary, rigid rotation are studied. In spite of the
seemingly non-Newtonian features of the bounding surface for some rotation rates,
we show, by means of a detailed analysis of the three-dimensional spatial geodesics,
that the standard Newtonian convexity properties do hold. A central role is played
by a family of geodesics that are introduced here, and provide a generalization of
the Newtonian straight lines parallel to the axis of rotation.
\end{abstract}

\vspace{0.5cm}

\hspace{0.2cm} PACS numbers: 04.20.Jb, 04.40.-b

\section{Introduction}
In the (thus far, unfulfilled) quest for a realistic exact solution 
in general relativity, representing
both the exterior and interior gravitational field generated by a self-gravitating 
axisymmetric mass of perfect fluid in stationary rotation, the detailed analysis
of the features of whatever partial results we already have seem relevant. 
Specifically, comparison with the known results in the Newtonian domain will 
improve our intuition within the general relativistic regime.

It is remarkable that there exists a number of treatments based on numerical 
integration of the field equations, or on approximation schemes valid for 
small rotation rates (applied, in particular, for the calculation of the shape of the
bounding surface of the fluid configurations, or for the analysis of the 
meaning of centrifugal forces), but, surprisingly, very few exact results, based
on the growing wealth of interior exact solutions in the literature (both
rigidly and differentially rotating). In the present paper, we analyze some
geometric features of one such interior solution, in order to check whether 
the analog of some Newtonian properties hold. Remarkably, they do, in spite 
of the fact that the analysis of the bounding surface  $p = 0$ in Section 3 naively
seems to point to the contrary. A more detailed analysis of the three-dimensional 
geodesics (Sections 4 and 5) shows that standard Newtonian features indeed hold
for the solution under consideration, if some of the Newtonian elements are
redefined appropriately. In particular, we introduce in Section 5 what we believe
is the generalization of straight lines parallel to the rotation axis in the
Newtonian case: Geodesics whose points have constant azimuthal angle and intersect
the equatorial plane orthogonally.

\section{The Kramer solution}
An exact solution of the Einstein field equations, representing the interior 
gravitational field of a self-gravitating, axially symmetric, rigidly rotating 
perfect fluid, was introduced in \cite{Kramer1}, and was further analyzed 
in \cite{Kramer2}. The metric can be written as
\begin{eqnarray}
   2 m d s^2 & = & [\eta - 1 - b \cos \xi e^{-\eta}] dt^2 + [4 (\eta 
   -1) - 4 b \cos \xi (e^{-\eta}-e^{-1})] dt d\varphi \label{eq:2.1} \\
   & & + [4 (\eta 
   -1) -4 b \cos \xi (e^{-\eta} + e^{\eta - 2} - 2 e^{-1})] d 
   \varphi^2 + \frac{d \eta^2}{\eta - 1} + \frac{e^{\eta}}{b \cos \xi} 
   d \xi^2\ ,\nonumber
\end{eqnarray}

\noindent where $m$ and $b$ are positive parameters. The coordinate $t$ is a time coordinate, 
while $\varphi$ is an azimuthal angle. The spacetime possesses the two commuting 
Killing fields  $\partial_{t}$       and  $\partial_{\varphi}$.           
 The axis of rotation is characterized 
by the equation $\eta = 1$, and there exists a discrete symmetry 
$\xi \rightarrow  -\xi$. The 
invariant set under this symmetry (i.e., points with $\xi = 0$) will be referred 
to as the {\it equatorial plane} in what follows, and the point with coordinates
$\eta = 1$, $\xi =0$ as the {\it center} of the body. The fluid obeys the following
barotropic equation of state

\begin{equation}
\varepsilon + 3 p = \frac{2 m}{\kappa_{0}}\ ,    
    \label{eq:2.2}
\end{equation}

\noindent where $p$ is the pressure and $\varepsilon$ the energy density, 
and $\kappa_{0}$ is a positive
constant. The dependence on $\eta$ and $\xi$
of the pressure and the energy density 
is the following:
\begin{eqnarray}
    p & = & {m \over 2 \kappa_{0}} (1 + \eta - b \cos \xi e^{-\eta})
    \label{eq:2.3}  \\
    \varepsilon & = & {m \over 2 \kappa_{0}} (1 - 3 \eta + 3 b \cos \xi 
    e^{-\eta})\ .
    \label{eq:2.4}
\end{eqnarray}

It is rather remarkable that the pressure is harmonic in  the $(\eta, 
\xi)$ coordinates:

\begin{equation}
    p_{\eta \eta} + p_{\xi \xi} = 0\ .
    \label{eq:2.5}
\end{equation}

\noindent Due to the minimum principle for the corresponding Laplacian, the pressure
attains its minimum value at the boundary of the domain of definition in 
the $(\eta, \xi)$ plane. This domain is given by the interior of the region
bounded by the line $\eta = 1$ and the curve $p(\eta , \xi)=0.$
As a matter of fact, the pressure has its lowest possible (negative) value at 
the center. The boundary value $p = 0$ (the greatest value) defines the boundary
of the object. In spite of the pressure being negative inside the body, and 
growing from the center to the boundary, the dominant energy condition is 
satisfied:

\begin{equation}
    \varepsilon > 0 , \ \ \  |p| < \varepsilon\ .
    \label{eq:2.6}
\end{equation}

It is remarkable that the boundary $p = 0$ has a relatively simple equation

\begin{equation}
   b \cos \xi - (1 + \eta ) e^{\eta} = 0\ . 
    \label{eq:2.7}
\end{equation}

\noindent This, and the fact that (as will be shown in Section 4) the integration 
of the relevant spatial geodesics can be reduced to quadratures, is crucial 
in our analysis of the geometric features of the solution.

The parameter $b$ is related to the modulus of the vorticity vector at the
center by means of

\begin{equation}
   (\omega_{\mu} \omega^{\mu})^{1 \over 2} = \sqrt{{m \over 2 b e} } (b + e)\ .
    \label{eq:2.8}
\end{equation}

When the requirement is made that the metric have the appropriate signature, 
as well as the requirement that  $\partial_{t}$  be timelike and  
$ \partial_{\varphi}$ spacelike, the
following inequalities result \cite{Kramer1}:

\begin{equation}
  m > 0, \ \ \eta \geq 1, \ \ b \cos \xi > 0, \ \ 1 - \eta + b \cos 
  \xi e^{-\eta} >0, \ \ (\eta + 1) (2 e^{\eta - 1} - e^{2 \eta -2}) 
  \geq 2 \ .
    \label{eq:2.9}
\end{equation}

We shall refer to the intersection of the equatorial plane with the boundary
$p = 0$ as the {\it equator} of the body, and to the region with $\xi > 0$ 
(respectively, $\xi < 0$) as the {\it northern} (resp., {\it southern}) 
{\it hemisphere}. Similarly, the intersection of the axis of rotation 
with the boundary $p = 0$ having $\xi > 0$ will be called the {\it north pole};
the intersection with$\ \xi < 0$ will be termed the {\it south pole}.
The object is oblate (in the sense that the polar distance to the center
is less than the distance from one point in the equator to the center), as
calculated in \cite{Kramer1}.

As $b$ is bounded away from zero, there is no static limit for this solution. 
One is tempted to interpret this feature in the light of the Newtonian result
\cite{Lamb} that the pressure cannot have a minimum at the center if 
$\Omega^2 < 2 \pi \rho$, where $\Omega$ is the angular velocity of 
the Newtonian fluid
body, and $\rho$ the mass density. It would be of interest to find the corresponding
result in general relativity.  

Finally, let us mention that the acceleration and the vorticity 
vectors are parallel at the pole, and orthogonal at the equator, as 
required for symmetry reasons.
\section{Geometry of the bounding surface $t=$const., $p=0$}

The spacetime metric  (\ref{eq:2.1}) induces the following metric 
on the two-dimensional
surface $t=$const., $p=0$:

\begin{eqnarray}
  2 m d s_2^{\ 2}  & = & g_{\varphi \varphi} d \varphi^2 + g_{\eta \eta} d \eta^2
     \label{eq:3.1}  \\
     & = & [4 (\eta + 1) (2 e^{\eta -1} - e^{2 \eta -2}) - 8] d 
     \varphi^2 + \frac{b^2 (\eta + 1) - (3 \eta + 5) e^{2 
     \eta}}{[b^2 -(1 + \eta)^2 e^{2 \eta}](\eta^2 -1)} d \eta^2\ ,
    \nonumber 
\end{eqnarray}

\noindent where we have used the equation for the surface $p=0$ (\ref{eq:2.7}) in order 
to express the
two-dimensional metric as a function of the coordinate $\eta$. It is remarkable
that (for large enough values of the rotation parameter $b$) the surface possesses
a region around the equator with {\it negative} Gaussian curvature $K$ 
(Figs. 1-2).
Our intuition with two-dimensional surfaces embedded in three-dimensional

\vspace{0.5cm} 

\noindent
\begin{minipage}[t]{.45 \linewidth}
 {\centering\epsfig{figure=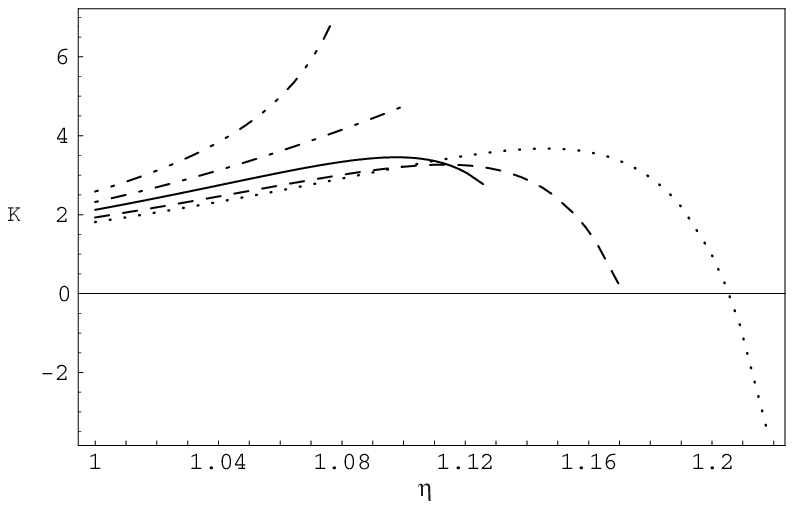,width=\linewidth}}
{\scriptsize {\bf Figure 1.} \ Gaussian curvature on the two-surface $ t =$ 
const., $p = 0$, from the pole ($\eta = 1$) to the equator, for different
values of the parameter $b$ ($b= 6.1, 6.3, 
6.5529, 7, 7.5$) corresponding, respectively, to the 
double-dot-dashed, dot-dashed, solid, dashed and  dotted lines in the figure. ($2m = 1$)} 
 \end{minipage} \hfill 
\begin{minipage}[t]{.50 \linewidth}
 {\centering\epsfig{figure=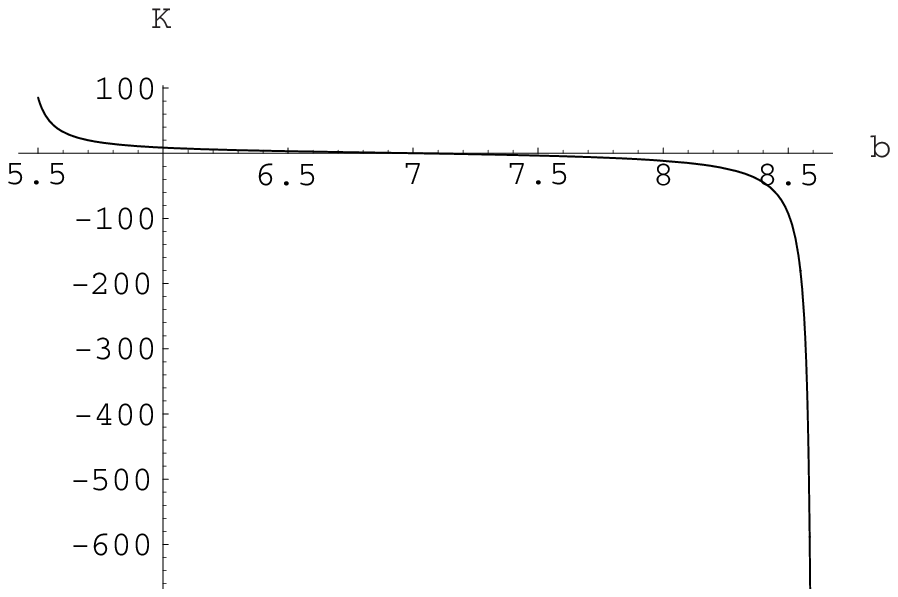,width=\linewidth}}
 
\vspace*{.5cm}

{\scriptsize {\bf Figure  2.} \ Gaussian curvature on the two-surface  $ t =$ 
const., $p = 0$, in the equator, as a function of $b$.}
 \end{minipage}  

\vspace{0.5cm}

\noindent Euclidean space would lead us to interpret that region as a
concave ``waist'' near the equator. This impression is reinforced by computing the
length of the parallels (closed curves of constant $\eta$, that circle around the
surface and can be parametrized by means of the azimuthal angle; they are not
geodesics, in general). The expression for the length of one such parallel,
obtained from  (\ref{eq:3.1}), is the following:

\begin{equation}
    l = \sqrt{{2 \over m}} \pi [4 (\eta + 1) (2 e^{\eta - 1} - e^{2 
    \eta - 2}) - 8 ]^{1 \over 2}\ .
    \label{eq:3.2}
\end{equation}

\noindent It is easily seen that the length presents a local maximum for

\begin{equation}
    2 \eta + 4 - (2 \eta + 3) e^{\eta -1} = 0\ .
    \label{eq:3.3}
\end{equation}

\noindent As a matter of fact, if the length is plotted as a function of  $\eta$  (Fig. 3),
we see  that it increases monotonically from   $\eta = 1$ (corresponding to the pole,
in which case the length vanishes) to a maximum at $\eta =1.1716$, obtained by
solving numerically (\ref{eq:3.3}). From that point on, the length decreases, 
until it
vanishes again for the numerical value  $\eta = 1.3134$. Please notice that the
preceding values, as well as (\ref{eq:3.3}) itself, do not depend on the rotation
parameter $b$. The dependence on $b$, however, shows up in the following: the
coordinate $\eta$ varies from   $\eta = 1$ (intersection of the rotation axis with the
surface $p=0$) to a maximum value (corresponding to the equator), obtained by
solving (\ref{eq:2.7}) with    $\xi =0$. 
Accordingly, the maximum value of  $\eta$   is an
increasing function of $b$. When $b < 7.0077$, the corresponding $\eta$  is such 
that it
falls within the left side of the curve in Fig. 3, and we have the ``normal''
situation, where the length of the parallels increases from one pole to the
equator. But, if $ b > 7.0077$, then the maximum length for a parallel occurs at an
intermediate latitude, and it subsequently decreases towards the equator. In the
extreme case $b = 8.603$ ($\eta  = 1.3134$), the circumferential length at the
equator vanishes, which could be interpreted as the fission of the body along the
equator at the extreme rotation rate. Due to positivity requirements in the
metric, the values   $\eta > 1.3134$ are excluded.

\begin{figure}
\centerline{\epsfig{figure=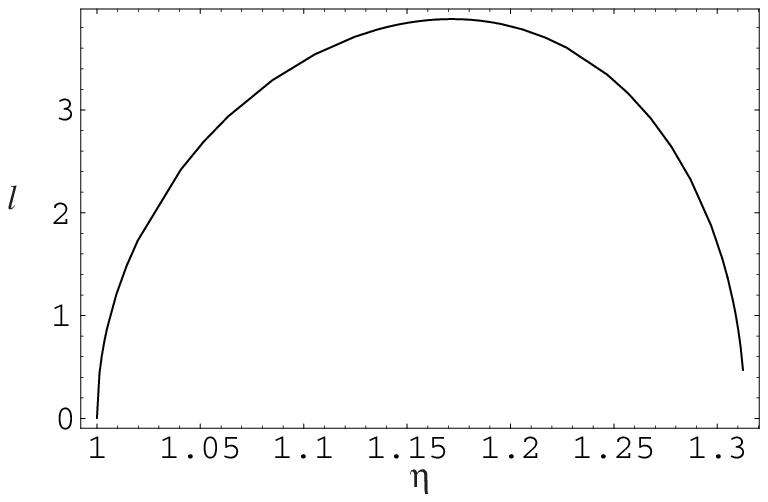,height=3in,
,bbllx=175bp,bblly=350bp,bburx=400bp,bbury=550bp,clip=}}
{\scriptsize {\bf Figure 3.} \ Length of the parallels (closed curves of $\eta =$ const.,
parametrized by $\varphi$ in the surface $p = 0$.)} 
\end{figure}

The Gaussian curvature of the two-surface vanishes precisely at  $\eta = 
1.1716$.
This is not accidental, as the curvature has a factor $g_{\varphi 
\varphi,\eta}$, and $g_{\varphi \varphi,\eta} = 0$ is
precisely the condition expressed by (\ref{eq:3.3}). Thus, the surface 
is ``flat'' at the
equator for $b = 7.0077$. If $b$ is increased, then a finite region with $K<0$ 
arises
symmetrically around the equator, including the equator itself. At the extreme
value $b = 8.603$, $K$ becomes singular (minus infinity, see Fig. 2) at the equator.

If the three-geometry where the two-surface $t =$ const., $p = 0$ is embedded were
Euclidean, we would find that the three-volume $t =$ const. enclosed by the
two-surface would not be convex, and, in particular, certain straight lines
parallel to the axis of rotation would intersect the boundary $p = 0$ in more than
two points (for sufficiently large values of $b$), against the well-known Newtonian
theorems of Lichtenstein, \cite{Lichtenstein1}, \cite{Lichtenstein2}, \cite{Lamb}.
We shall see below, however, that a natural generalization
of the mentioned parallel straight
lines to the true three-dimensional Riemannian geometry does preserve the analog
of the classical results.

It should be remarked that the closed curves  $\eta=$ const. on the boundary surface
are not geodesics on the surface, except for the particular case where 
$g_{\varphi \varphi,\eta} = 0$. This can be readily 
seen by writing the
equations for the geodesics in the metric (\ref{eq:3.1})

\begin{eqnarray}
    g_{\varphi \varphi} \dot{\varphi} & = & \mathrm{const.}
    \label{eq:3.4}  \\
    2 g_{\eta \eta} \ddot{\eta} + g_{\eta \eta ,\eta} \dot{\eta}^2 - 
    g_{\varphi \varphi, \eta} \dot{\varphi}^2 & = & 0
    \label{eq:3.5}
\end{eqnarray}
(where a dot denotes a derivative with respect to length along the 
geodesic).

We thus see that the only parallel circles which are geodesics are the two
(symmetrically placed with respect to the equator) corresponding to  
$\eta  = 1.1716$, when $b > 7.0077$. When $b = 7.0077$, the two parallels 
coincide with
the equator (and that is the only case in which the equator is a geodesic).

Finally, it is easily shown (by numerically computing the Gauss-Bonnet integral
of $K$ over the surface) that the two-surface $t =$ const., $p = 0$ has 
the topology of a 2-sphere.

\section{Geodesics of the three-dimensional spatial\\ 
metric}

The spacetime metric (\ref{eq:2.1}) reduces for $t =$ const. to the following
three-dimensional metric:

\begin{equation}
    2 m d s_{3}^{\ 2} = 4 [ \eta - 1 - b \cos \xi (e^{-\eta} + 
    e^{\eta-2}-2 e^{-1})] d \varphi^2 + \frac{d \eta ^2}{\eta - 1} + 
    \frac{e^{\eta}}{b \cos \xi} d \xi^2\ .
    \label{eq:4.1}
\end{equation}

From (\ref{eq:4.1}), we find the following equations for the geodesics in the
three-space:

\begin{eqnarray}
    [\eta - 1 - b \cos \xi (e^{-\eta} +e^{\eta -2} - 2 e^{-1})] 
    \dot{\varphi} = \mathrm{const.} &  &
    \label{eq:4.2}  \\
    \frac{2 \ddot{\eta}}{\eta -1} - \frac{\dot{\eta}^2}{(\eta 
    -1)^2} -\frac{e^{\eta}}{b \cos \xi} \dot{\xi}^2 - 4 [1 - b \cos 
    \xi (-e^{-\eta} + e^{\eta -2})]\dot{\varphi}^2 & = & 0
    \label{eq:4.3}  \\
    \frac{2 e^{\eta}}{b \cos \xi} \ddot{\xi} + \frac{2 e^{\eta} 
    \dot{\eta}\dot{\xi}}{b \cos\xi} + \frac{e^{\eta} \sin \xi}{b 
    \cos^2\xi} \dot{\xi}^2 - 4 b \sin \xi 
    (e^{-\eta} +e^{\eta -2} - 2 e^{-1}) \dot{\varphi}^2 & =  & 0.
    \label{eq:4.4}
\end{eqnarray}

In particular, geodesics with $\dot{\varphi}  = 0$ are characterized by 
the two equations

\begin{eqnarray}
    \frac{2 \ddot{\eta}}{\eta -1} - \frac{\dot{\eta}^2}{(\eta 
    -1)^2} -\frac{e^{\eta}}{b \cos \xi} \dot{\xi}^2 & = & 0
    \label{eq:4.5}  \\
    \ddot{\xi} + \dot{\eta} \dot{\xi} + {1 \over 2} \frac{\sin 
    \xi}{\cos \xi} \dot{\xi}^2 & = & 0\ .
    \label{eq:4.6}
\end{eqnarray}

It is a rather remarkable feature of the Kramer solution that the geodesic
equations (\ref{eq:4.5})-(\ref{eq:4.6}) can be reduced to quadratures. 
Two clearly different cases
appear, depending on whether  $\dot{\xi} = 0$ at all points of the geodesic 
or not. In the former case, the integration reduces to that of the equation

\begin{equation}
 \frac{2 \ddot{\eta}}{\eta -1} - \frac{\dot{\eta}^2}{(\eta 
    -1)^2} = 0\ , 
    \label{eq:4.7}
\end{equation}
which yields
\begin{equation}
    s = q \sqrt{{2 \over m}} [ \sqrt{\eta_{f}-1} - \sqrt{\eta_{i} - 
    1}]\ ,
    \label{eq:4.8}
\end{equation}
where $q = \pm 1$, depending on the sense in which the geodesic is 
traversed; $\eta_{i}$ is the initial $\eta$ coordinate, and 
$\eta_{f}$ the final one, and 
$s$ is the distance along the geodesic. In the case   $\dot{\xi} \neq 0$, 
we introduce the
new variable $w = \sqrt{\eta -1}$, in order to simplify the equations. 
By dividing eq. (\ref{eq:4.6})   
by $\dot{\xi}$, it can be immediately integrated once, giving

\begin{equation}
    \frac{\dot{\xi}^2}{\cos \xi} = k e^{- 2 \eta}\ ,
    \label{eq:4.9}
\end{equation}
where $k$ is a positive constant. By substituting (\ref{eq:4.9}) into 
(\ref{eq:4.5}), we get
\begin{equation}
    \ddot{w} - \frac{k}{4 b e} w e^{-w^2} = 0\ .
    \label{eq:4.10}
\end{equation}
If $\dot{w} = 0$, we get the system
\begin{eqnarray}
    w & = & 0
    \label{eq:4.11º}  \\
    \frac{\dot{\xi}^2}{\cos \xi} & = &  k e^{-2}\ .
    \label{eq:4.12}
\end{eqnarray}

\noindent This corresponds to the geodesic along the rotation axis. In the generic 
case, $ \dot{w} \neq 0$, upon multiplication of (\ref{eq:4.10})  by 
$\dot{w}$ we get

\begin{equation}
    \ddot{w} \dot{w} - \frac{k}{4 b e} e^{-w^2} w \dot{w} = 0\ ,
    \label{eq:4.13}
\end{equation}
whose first integral is
\begin{equation}
    4 \dot{w}^2 + \frac{k}{be} e^{-w^2} = \alpha
    \label{eq:4.14}
\end{equation}
with $\alpha > 0$ a constant of integration. One can now express the relation among
the coordinate $w$ on the geodesic and the distance $s$ along the geodesic by
\begin{equation}
  \frac{2  d w}{\sqrt{\alpha - \frac{k}{be} e^{-w^2}}} = q d s 
    \label{eq:4.15}
\end{equation}
($q = \pm 1$), while the equation for the trajectory is given by
\begin{equation}
    \frac{d \xi}{\sqrt{\cos \xi}} = 2 \epsilon q { \sqrt{k} \over 
    e} \frac{e^{-w^2}}{\sqrt{\alpha - {k \over be} e^{-w^2}}} d w\ ,
    \label{eq:4.16}
\end{equation}
where $\epsilon = \pm 1$. By using (\ref{eq:4.1}), we find
$\alpha = 2m$. To summarize, the relevant equations can we written as
\begin{eqnarray}
    \dot{w} & = & q \sqrt{{m \over 2}} {1 \over \sqrt{\beta}} 
    \sqrt{\beta - e^{-w^2}}
    \label{eq:4.17}  \\
    \dot{\xi} & = & 2 \epsilon q \sqrt{{m \over 2}} {1 \over \sqrt{\beta}}
    \sqrt{{b \over e}} \sqrt{\cos \xi} e^{-w^2}
    \label{eq:4.18}
\end{eqnarray}
where $\beta = {b e \alpha \over k}$.

Eqs. (\ref{eq:4.17}) and (\ref{eq:4.18}) can be expressed as quadratures:
\begin{eqnarray}
    \frac{d \xi}{\sqrt{\cos \xi}} & = & 2 \epsilon \sqrt{b \over 
    e} \frac{e^{-w^2}}{\sqrt{\beta - e^{-w^2}}} d w
    \label{eq:4.19}  \\
    d s & = & q \sqrt{2 \over m} \sqrt{\beta} \frac{d w}{\sqrt{\beta - e^{-w^2}}}
    \label{eq:4.20}
\end{eqnarray}
(Notice that $q$ and $\epsilon$ are signs, which can be chosen so that the 
distance $s$ along
the geodesic increases from the initial value $s = 0$ at the initial point.)

\section{Convexity properties of the fluid body}

Given the results in Section 3, one could naively expect that the distance from
points in the surface $p = 0$ to the axis of rotation would decrease (for large
enough values of $b$) as the equator is approached. We shall see that this is not
the case. In order to do that, let us first identify some general properties of
geodesics starting from some point in the symmetry axis and reaching a point in
the boundary $p = 0$. For definiteness, we shall work in the northern hemisphere;
due to the symmetry with respect to the equatorial plane, 
analogous considerations hold
for the southern hemisphere.

The first observation is that geodesics from the axis of rotation lie in the
$(w,\xi)$ plane, with constant $\varphi$. This can be seen from eq. (\ref{eq:4.2}): 
If the value $w = 0 \ (\eta = 1)$, 
characterizing the axis, is substituted, then the left-hand side
vanishes. Therefore, the constant on the right vanishes. That shows that 
$\dot{\varphi}  = 0$ for
such geodesics. The relevant equations for the geodesics are then
(\ref{eq:4.5})-(\ref{eq:4.6}), whose
integrals are given by (\ref{eq:4.19})-(\ref{eq:4.20}). Next, we find the 
meaning of the constant
$\beta$: It is easily seen, by using the standard Riemannian formula, that the angle
$\gamma$ between the axis and the tangent to the geodesic at the axis (defined such
that $\gamma = 0$ for a geodesic starting at the axis and pointing towards the north
pole) is related to $\beta$ by the following relation:

\begin{equation}
  \label{eq:5.1}
  \cos \gamma = {1 \over \sqrt{\beta}}\ .
\end{equation}

For a given point in the boundary $p = 0$, characterized by $w = w_{f}$, it is found
that a geodesic joining it to the axis has a distance $s$ to the axis given by
\begin{equation}
  \label{eq:5.2}
  s = \sqrt{{2 \over m}} \int_{0}^{w_{f}} \frac{d w}{\sqrt{1-\cos^2 \gamma e^{-w^2}}}\ .
\end{equation}

From (\ref{eq:5.2}), we see that
\begin{equation}
  \label{eq:5.3}
  \frac{\partial s}{\partial \gamma} = - \sqrt{{2 \over m}} \sin \gamma
  \cos \gamma \int_{0}^{w_{f}} \frac{e^{-w^2}}{(1-\cos^2\gamma e^{-w^2})^{3 \over 2}} d w\ .
\end{equation}

\noindent As a consequence, $s$ is a decreasing function of $\gamma$ in the northern hemisphere.
The minimum obtains for $\gamma = {\pi \over 2}$, which corresponds to a geodesic 
with constant $\xi$, whose length (\ref{eq:4.8}) is
\begin{equation}
  \label{eq:5.4}
  s = \sqrt{{2 \over m}} w_{f}\ .
\end{equation}

\noindent This, being the minimum of the different distances along different geodesics to
the axis, we will define as {\it the distance to the axis} from the given point
$(w_{f}, \xi_{f})$ in the boundary. Let us now look at how the distance so
defined varies when we consider different points in the boundary. By using the
equation for the boundary,
\begin{equation}
  \label{eq:5.5}
  (w_{f}^{\ 2} + 2) e^{w_{f}^{\ 2}} = {b \over e} \cos \xi_{f}\ ,
\end{equation}

\noindent and the derivative of $w_{f}$ with respect to $\xi_{f}$ along the boundary,

\begin{equation}
  \label{eq:5.6}
  \frac{d w_{f}}{d \xi_{f}} = - \left({b \over e} \right) \frac{e^{-w_{f}^{\ 2}}}
  {2 w_{f}^{\ 3} + 6 w_{f}} \cos \xi_{f} \ ,
\end{equation}

\noindent we get
\begin{equation}
  \label{eq:5.7}
  \frac{\partial s}{\partial \xi_{f}} = \frac{\partial s}{\partial w_{f}}
  \frac{d w_{f}}{d \xi_{f}} = - \sqrt{{2 \over m}} \left({b \over e}
  \right) \frac{e^{-w_{f}^{\ 2}}}{2 w_{f}^{\ 3} + 6 w_{f} }\cos\xi_{f}\ ,
\end{equation}

\noindent thus showing that the distance from a point in the boundary to the axis {\it
increases} monotonically from the north pole to the equator.

Another measure of the convexity (or lack thereof) of the boundary $p = 0$ is the
behaviour of the distances from the center $(w = 0, \xi = 0)$ to points in the
boundary (Fig. 4). Let us denote by $w_{f}(\gamma)$ the $w$ coordinate of the 
endpoint of a geodesic starting from the center with an angle $\gamma$ with 
respect to the northern semiaxis. The distance to the endpoint will be given by
\begin{equation}
  \label{eq:5.8}
  \sqrt{{m \over 2}} s = \int_{0}^{w_{f}(\gamma)} \frac{1}{\sqrt{1-\cos^2\gamma e^{-w^2}}} d w \ .
\end{equation}

\begin{figure}
\centerline{\epsfig{figure=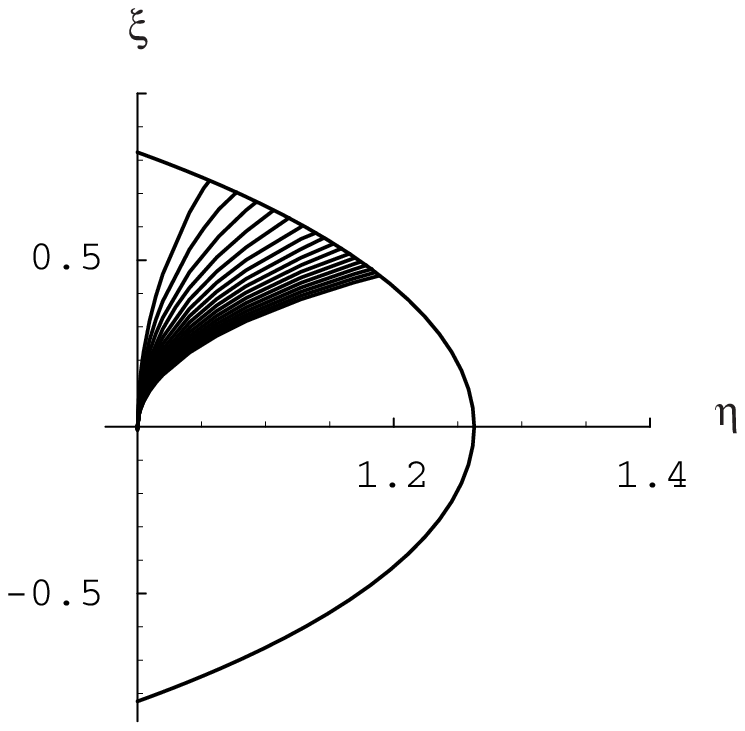,height=3in,
,bbllx=50bp,bblly=300bp,bburx=400bp,bbury=550bp,clip=}}
{\scriptsize {\bf Figure 4.} \ Trajectories of geodesics $\varphi=$ const. from the center to
the surface $p = 0$, with different values of initial velocities, i.e., 
varying $\beta$}
\end{figure}

According to the previous equation, the derivative of $s$ with respect to $\gamma$ is
\begin{equation}
  \label{eq:5.9}
  \sqrt{{m \over 2}} \frac{\partial s}{\partial \gamma} = \frac{\partial w_{f}(\gamma)}{\partial 
\gamma} \frac{1}{\sqrt{1-\cos^2\gamma e^{-w_{f}^{\ 2}(\gamma)}}} - 
  \int_{0}^{w_{f}(\gamma)} \frac{\sin\gamma \cos\gamma e^{-w^2}}{(1-\cos^2\gamma 
  e^{-w^2})^{3 \over 2}} d w\ .
\end{equation}

\noindent We are interested in the growth properties of $s$ at the equator. In order to
evaluate the derivative $\frac{\partial w_{f}(\gamma)}{\partial \gamma}$, we consider 
the equation for the boundary (\ref{eq:5.5}), and differentiate it with respect to $\gamma$:
\begin{equation}
  \label{eq:5.10}
  - {b \over e} \sin\xi_{f}(\gamma) \frac{\partial \xi_{f}(\gamma)}{\partial \gamma} = 
  \left[2 w_{f}^{\ 3}(\gamma) + 6 w_{f}(\gamma)\right] e^{w_{f}^{\ 2}(\gamma)} \frac{\partial w_{f}
  (\gamma)}{\partial \gamma}\ .
\end{equation}

\noindent The geodesic with $\gamma = {\pi \over 2}$ corresponds to a geodesic along the equatorial
plane, with $\xi = 0$. Hence, $\xi_{f}({\pi \over 2}) = 0$. Substituting $\gamma = 
{\pi \over 2}$ in (\ref{eq:5.10}), one gets
\begin{equation}
  \label{eq:5.11}
  \left[2 w_{f}^{\ 3}\left({\pi \over 2}\right) + 6 w_{f}\left({\pi \over 2}\right)\right]
 e^{w_{f}^{\ 2}({\pi \over 2})} 
  \left.\frac{\partial w_{f}(\gamma)}{\partial \gamma} \right|_{\gamma = {\pi \over 2}} = 0\ .
\end{equation}

\noindent But the coefficient in front of the derivative does not vanish; therefore,
\begin{equation}
  \label{eq:5.12}
  \left.\frac{\partial w_{f}(\gamma)}{\partial \gamma} \right|_{\gamma = {\pi \over 2}} = 0\ .
\end{equation}

\noindent By substituting this result in eq. (\ref{eq:5.9}), evaluated at $\gamma = 
{\pi \over 2}$, we get
\begin{equation}
  \label{eq:5.13}
  \left.\frac{\partial s}{\partial \gamma} \right|_{\gamma = {\pi \over 2}} = 0\ ,
\end{equation}

\noindent thus showing that the distance from the center to the equator is a local
extremum, if the second derivative of $s$ with respect to $\gamma$ does not vanish.
Its sign will then decide whether the equatorial distance is a maximum
(corresponding to local convexity at the equator) or a minimum (local concavity).
The second derivative of $s$ with respect to $\gamma$ will be given from eq. (\ref{eq:5.9})
by
\begin{eqnarray}
    \sqrt{{m\over2}} \frac{\partial^2 s}{\partial \gamma^2} & = & 
    \frac{\partial^2 w_{f}(\gamma)}{\partial \gamma^2} 
    (1 - \cos^2 \gamma e^{-w_{f}^{\ 2}(\gamma)})^{-{1 \over 2}}-(1 -\cos^2 \gamma 
    e^{-w_{f}^{\ 2}(\gamma)})^{-{3 \over 2}} \times \nonumber \\
   && [\cos \gamma \sin \gamma  + \cos^2 
    \gamma w_{f}(\gamma)]e^{-w_{f}^{\ 2}(\gamma)}\left[\frac{\partial 
    w_{f}(\gamma)}{\partial \gamma}\right]^2 \nonumber \\
    && - \sin \gamma \cos \gamma 
    e^{-w_{f}^{\ 2}(\gamma)} (1 -\cos^2 \gamma 
    e^{-w_{f}^{\ 2}(\gamma)})^{-{3 \over 2}} \frac{\partial 
    w_{f}(\gamma)}{\partial \gamma}   \nonumber \\ 
    && + (\sin^2 \gamma - \cos^2 \gamma )
    \int_{0}^{w_{f}(\gamma)} \frac{e^{-w^2}}{(1 - \cos^2 \gamma 
    e^{-w^2})^{3 \over 2}} d w \nonumber \\
   && -3 \sin^2 \gamma \cos^2 \gamma 
    \int_{0}^{w_{f}(\gamma)} \frac{e^{-2 w^2}}{(1 - \cos^2 \gamma 
    e^{-w^2})^{5 \over 2}} d w \ .
    \label{eq:5.14}
\end{eqnarray}

\noindent At $\gamma = {\pi \over 2}$, the preceding equation reduces to
\begin{equation}
    \left.\sqrt{m \over 2} \frac{\partial^2 s}{\partial 
    \gamma^2}\right|_{\gamma={\pi \over 2}} = 
     \left.\frac{\partial^2 w_{f}(\gamma)}{\partial 
    \gamma^2}\right|_{\gamma={\pi \over 2}} + \int_0^{w_{f}({\pi \over 2})}e^{-w^2} d w\ .
    \label{eq:5.15}
\end{equation}

\noindent In order to evaluate the second derivative $\left.\frac{\partial^2 w_{f}
(\gamma)}{\partial \gamma^2}\right|_{\gamma={\pi \over 2}}$, we first
consider the equation for the trajectory of a geodesic from the center;
from (\ref{eq:4.19}),

\begin{equation}
    \int_{0}^{\xi_{f}(\gamma)} \frac{d \xi}{\sqrt{\cos \xi}} = 2 
    \sqrt{b \over e} \cos \gamma 
    \int_{0}^{w_{f}(\gamma)} \frac{e^{-w^2}}{\sqrt{1 - \cos^2 \gamma 
    e^{-w^2}}} d w\ .
    \label{eq:5.16}
\end{equation}

\noindent By differentiating (\ref{eq:5.16}) with respect to $\gamma$, we find

\begin{eqnarray}
    \frac{\partial \xi_{f}(\gamma)}{\partial \gamma} {1 \over 
    \sqrt{\cos \xi_{f}(\gamma)}} & = & 2 \sqrt{b \over e} \cos \gamma 
     \frac{\partial w_{f}(\gamma)}{\partial \gamma}
     \frac{e^{-w_{f}^{\ 2}(\gamma)}}{\sqrt{1 - \cos^2 \gamma 
    e^{-w_{f}^{\ 2}(\gamma)}}}  \nonumber \\
    & & - 2 \sqrt{b \over e} \sin \gamma 
    \int_{0}^{w_{f}(\gamma)} \frac{e^{-w^2} d w}{(1 - \cos^2 \gamma 
    e^{-w^2})^{3 \over 2}}\ .  
    \label{eq:5.17}
\end{eqnarray}

\noindent Hence

\begin{equation}
    \left. \frac{\partial \xi_{f}(\gamma)}{\partial \gamma} 
    \right|_{\gamma= {\pi \over 2}} = -2 \sqrt{b \over e} 
    \int_{0}^{w_{f}({\pi \over 2})} e ^{-w^2} d w\ .
    \label{eq:5.18}
\end{equation}

\noindent By differentiating (\ref{eq:5.10}) we obtain

\begin{eqnarray}
    &&-{b \over e} \cos \xi_{f}(\gamma) \left[ \frac{\partial 
    \xi_{f}(\gamma)}{\partial \gamma}\right]^2 - {b \over e} \sin 
    \xi_{f}(\gamma) \frac{\partial^2 
    \xi_{f}(\gamma)}{\partial \gamma^2}  \label{eq:5.19} \\
    &&= \frac{\partial}{\partial w_{f}(\gamma)} \left\{
    [2 w_{f}^{\ 3}(\gamma) + 6  w_{f}(\gamma)] e^{ w_{f}^{\ 2}(\gamma)}
    \right\}
    \left[ \frac{\partial 
    w_{f}(\gamma)}{\partial \gamma}\right]^2 \nonumber \\
    &&+ [2 w_{f}^{\ 3}(\gamma) + 6 
    w_{f}(\gamma)]e^{w_{f}^{\ 2}(\gamma)}
    \frac{\partial^2  
    w_{f}(\gamma)}{\partial \gamma^2} \ ,
     \nonumber
\end{eqnarray}

\noindent and, setting $\gamma = {\pi \over 2}$ in (\ref{eq:5.19}),

\begin{equation}
    -{b \over e}  \left[ \left.\frac{\partial 
    \xi_{f}(\gamma)}{\partial 
    \gamma}\right|_{\gamma={\pi \over 2}}\right]^2 = 
   \left[2 w_{f}^{\ 3}\left({\pi \over 2}\right) + 6 
    w_{f}\left({\pi \over 2}\right)\right] e^{w_{f}^{\ 2}\left({\pi \over 2}\right)}
    \left.\frac{\partial^2  
    w_{f}(\gamma)}{\partial \gamma^2}\right|_{\gamma={\pi \over 2}} \ .
    \label{eq:5.20}
\end{equation}

\noindent Finally, from (\ref{eq:5.20}) and (\ref{eq:5.18}) we get

\begin{equation}
  \left.\frac{\partial^2  
    w_{f}(\gamma)}{\partial \gamma^2}\right|_{\gamma={\pi \over 2}} = - 2 
    \left({b \over e}\right)^2 \frac{e^{-w_{f}^{\ 2}({\pi \over 2})}}{w_{f}^{\ 3}
    ({\pi \over 2}) + 3 
    w_{f}({\pi \over 2})} \left[
    \int_{0}^{w_{f}({\pi \over 2})} e^{-w^2} d w
    \right]^2\ .
    \label{eq:5.21}
\end{equation}

\noindent Going back to (\ref{eq:5.15}), and substituting $\left.\frac{\partial^2 w_{f}
(\gamma)}{\partial \gamma^2}\right|_{\gamma={\pi \over 2}}$ from (\ref{eq:5.21}), the
following expression for the second derivative of the distance is obtained:

\begin{equation}
    \sqrt{m \over 2} \left.\frac{\partial^2 s}{\partial \gamma^2} 
    \right|_{\gamma={\pi \over 2}} = \int_{0}^{w_{f}({\pi \over 2})}e^{-w^2} d w 
    \left[
    1 - \frac{2 
    ({b \over e})^2 e^{-w_{f}^{\ 2}({\pi \over 2})}}{w_{f}^{\ 3}({\pi \over 2}) + 3 
    w_{f}({\pi \over 2})} \int_{0}^{w_{f}({\pi \over 2})} e^{-w^2} d w
    \right]\ .
    \label{eq:5.22}
\end{equation}

\noindent But, substituting ${b \over e}$ from (\ref{eq:5.5}) (with $\xi_{f} = 0)$
in (\ref{eq:5.22}), we find the following inequality:

\begin{eqnarray}
    \frac{2 
    ({b \over e})^2 e^{-w_{f}^{\ 2}({\pi \over 2})}}{w_{f}^{\ 3}({\pi \over 2}) + 3 
    w_{f}({\pi \over 2})} \int_{0}^{w_{f}({\pi \over 2})} e^{-w^2} d w  =  
    \frac{2 [ w_{f}^{\ 2}({\pi \over 2}) + 2]^2
     e^{w_{f}^{\ 2}({\pi \over 2})}}{w_{f}^{\ 3}({\pi \over 2}) + 3 
    w_{f}({\pi \over 2})} \int_{0}^{w_{f}({\pi \over 2})} e^{-w^2} d w &&
   \nonumber  \\
     \geq \frac{2 [ w_{f}^{\ 2}({\pi \over 2}) + 2]^2
     e^{-w_{f}^{\ 2}({\pi \over 2})}}{w_{f}^{\ 3}({\pi \over 2}) + 3 
    w_{f}({\pi \over 2})} \int_{0}^{w_{f}({\pi \over 2})} e^{-w_{f}^{\ 2}({\pi \over 2})} d w  =  
    \frac{2 [ w_{f}^{\ 2}({\pi \over 2}) + 2]^2}{w_{f}^{\ 2}({\pi \over 2}) + 3} > 2\ , \ \ \ \ &&
    \label{eq:5.23}
\end{eqnarray}

\noindent thus showing that
\begin{equation}
  \label{eq:5.24}
  \left.\frac{\partial^2 s}{\partial \gamma^2}\right|_{\gamma={\pi \over 2}} < 0\ .
\end{equation}

It should be stressed that (\ref{eq:5.24}) does not depend on $b$. We conclude that the
distance from the center presents a local {\it maximum} at the equator, thus
showing that our naive expectations from the analysis in Section 3 were
unfounded.

Let us now consider geodesics joining points symmetrically placed with respect to
the equator, $(w_{0}, \xi_{0})$ and $(w_{0}, -\xi_{0})$, and having $\dot{\varphi} = 0$.
For symmetry and differentiability reasons, such geodesics must intersect 
the equatorial plane orthogonally, with respect to the metric

\begin{equation}
  \label{eq:5.25}
  {2 \over m} d w^{2} + \frac{1}{2 m} \left({e \over b}\right) \frac{e^{w^2}}{\cos\xi} d \xi^2\ .
\end{equation}

The orthogonality condition fixes the parameter $\beta$ in (\ref{eq:4.19}) and (\ref{eq:4.20}):
 
\begin{equation}
  \label{eq:5.26}
  \beta = e^{-w_{c}^{2}}\ ,
\end{equation}

\noindent where $w_c$ is the $w$ coordinate of the intersection of the geodesic with the
equatorial plane. We shall now consider the portion of the geodesic starting
orthogonally to the equatorial plane from $(w_c,0)$ and ending in $(w_0,\xi_0)$, whose length
will obviously be half that of the complete geodesic starting at $(w_0, -\xi_0)$ and
ending in $(w_0,\xi_0)$. The signs in (\ref{eq:4.17}) and (\ref{eq:4.20})  are fixed 
by the initial conditions, giving
\begin{equation}
  \label{eq:5.27}
  \epsilon q = + 1\ .
\end{equation}

In principle, the considered portion of geodesic in the northern hemisphere could
have $\dot{w} > 0$ or $\dot{w} < 0$. But the latter does not, in fact, exist, as one
would have
\begin{equation}
  \label{eq:5.28}
  \beta - e^{-w^2} < 0\ ,
\end{equation}

\noindent due to the fact that $w_0 < w_c$ for  $\dot{w} < 0$: From (\ref{eq:4.19}), such 
a possibility is
incompatible with the equation for the geodesic trajectory. We conclude that the
unique geodesic joining $(w_0, -\xi_0)$ and $(w_0,\xi_0)$ cuts orthogonally the
equatorial plane at $(w_c, 0)$, with $w_c < w_0,$ and exhibits the monotonic 
behaviour $\dot{w} > 0$ and  $\dot{\xi} > 0$ 
[from (\ref{eq:5.27}) and (\ref{eq:4.17})] in the northern hemisphere.

Such a geodesic is the natural generalization of a straight line parallel to the
rotation axis in the Newtonian case: Both can be defined as non-twisting 
$(\dot{\varphi} = 0)$ geodesics that intersect the equatorial plane orthogonally. We shall now
show that the geodesic thus introduced is completely contained in the
three-dimensional body bounded by the surface $t =$ const., $p = 0.$ To this end, we
consider the pressure $p$ as a function of a point in the geodesic; from (\ref{eq:2.3}),

\begin{equation}
  \label{eq:5.29}
  2\frac{\kappa_{0}}{m} p(s) = w^2(s) + 2 - {b \over e} \cos\xi(s) e^{-w^2(s)}\ .
\end{equation}

\noindent By differentiating (\ref{eq:5.29}) with respect to the distances from the starting point
$(w_c,0)$ along the portion of the geodesic in the northern hemisphere (we denote
the derivative by  a dot), we find

\begin{equation}
  \label{eq:5.30}
  2 {\kappa_{0} \over m} \dot{p} = (2 w + 2 {b \over e} \cos\xi w e^{-w^2}) \dot{w} +
  {b \over e} \sin\xi e^{-w^2} \dot{\xi}\ ;
\end{equation}

\noindent but, due to the inequalities  $\dot{w} > 0$ and  $\dot{\xi} > 0$ for the northern 
portion of the geodesic, and the fact that $2 w + 2 {b \over e} \cos\xi w e^{-w^2} > 0$ 
and ${b \over e} \sin\xi e^{-w^2} > 0,$ we conclude that
\begin{equation}
  \label{eq:5.31}
  \dot{p} > 0\ .
\end{equation}

\noindent As a consequence, the pressure along the geodesic increases from $(w_c, 0)$ to the
endpoint $(w_0, \xi_0)$. Conversely, traversing the geodesic in the opposite sense
[starting from $(w_0, \xi_0)$ and heading towards $(w_c, 0)$] corresponds to
decreasing values of $p$. As the pressure decreases towards the interior in the
solution under consideration, it is clear that a geodesic of the type just
introduced which starts at a point in the surface $(p = 0)$ in the northern
hemisphere has $p < 0$ at all other points in the northern hemisphere. By symmetry,
all other points of the geodesic in the southern hemisphere have $p < 0$, except
for the final point, where $p = 0.$ Thus, a geodesic of the type considered has
only two points of intersection with the bounding surface, thus maintaining in
the present fluid configuration the classical (Newtonian) result for the
intersection of straight lines parallel to the rotation axis with the boundary $p
= 0.$

We have found that all the criteria we have considered reproduce the convexity
properties of Newtonian configurations, in spite of the peculiar behaviour of the
boundary surface, as analyzed in Section 3. It is clear that the standard
Euclidean relations among convex bodies and their bounding surfaces, \cite{Hadamard},
\cite{Bonnesen}, \cite{Chern}, do not hold for a general Riemannian geometry, dynamically
prescribed by Einstein's equations.

\section{Conclusions}

We have analyzed some geometric features related to the shape and convexity
properties of a self-gravitating body of perfect fluid in stationary rotation,
as given by the exact solution \cite{Kramer1}. Our analysis of the 
boundary $(t=$ const., $p = 0)$ shows that for some rotation rates there appear features
that would be interpreted in the Euclidean case as non-Newtonian. However, by looking at
the behaviour of geodesics in the three-dimensional fluid within the 
bounding surface, the analog of the Newtonian results under consideration is 
obtained. The technique to show them is a detailed analysis of the spatial geodesics
within the fluid, and specifically the introduction of a family of geodesics
which generalize the straight lines parallel to the axis of rotation in the Newtonian
case. 
\section{Acknowledgments}

We thank L. Fern\'andez-Jambrina, L. M. Gonz\'alez-Romero, and F. Navarro
for discussions. Financial support from Direcci\'on General de Ense\~nanza
Superior e Investigaci\'on Cient\'{\i}fica (Project PB95-0371) is gratefully
acknowledged.


\end{document}